\documentclass[twocolumn,conference,dvipdfmx]{IEEEtran}

\usepackage[dvipdfmx]{graphicx}

\usepackage{amssymb,amsfonts,amstext}
\usepackage{amsmath}	
\usepackage{latexsym}
\usepackage{color}

\newcommand{\be}{\begin{equation}}
\newcommand{\ee}{\end{equation}}

\newtheorem{definition}{Definition}

\newtheorem{theorem}[definition]{Theorem}

\def\QED{\mbox{\rule[0pt]{1.5ex}{1.5ex}}}
\def\endproof{\hspace*{\fill}~\QED\par\endtrivlist\unskip}

\begin{document}
\title{Non-Malleable Codes Against Affine Errors}

\author{\IEEEauthorblockN{Ryota Iwamoto}\IEEEauthorblockA{Graduate School of Science and Engineering\\
Saitama Univeristy\\ Email: s13ti006@mail.saitama-u.ac.jp}\and
\IEEEauthorblockN{Takeshi Koshiba}\IEEEauthorblockA{Graduate School of Science and Engineering\\
Saitama University\\ Email: koshiba@mail.saitama-u.ac.jp}}

\maketitle

\begin{abstract}
Non-malleable code is a relaxed version of error-correction codes
and the decoding of modified codewords results in the original message
or a completely unrelated value. Thus, if an adversary corrupts a codeword
then he cannot get any information from the codeword. This means that
non-malleable codes are useful to provide a security guarantee 
in such situations 
that the adversary can overwrite the encoded message.
In 2010, Dziembowski et al.\ showed a construction for non-malleable codes
against the adversary who can falsify codewords bitwise independently.
In this paper, we consider an extended adversarial model (affine error model)
where
the adversary can falsify codewords bitwise independently
or replace some bit with the value obtained by applying an affine map
over a limited number of bits. We prove that the non-malleable codes
(for the bitwise error model) provided by Dziembowski et al.\ are 
still non-malleable against the adversary in the affine error model.
\end{abstract}

\section{Introduction}
As we know, error-correction codes can recover
the original message from a corrupted codeword (within admissible errors)
and error-detection codes can detect if the codeword is corrupted 
while the error correction may not be possible.
The notion of {\em non-malleable codes},
invented by Dziembowski, Pietrzak, and Wichs \cite{1},
is a relaxed notion of error detection codes or error correction codes.
(The term ``non-malleability'' comes from non-malleable cryptography \cite{8}.
For non-malleable codes, we suppose that errors would be caused
by some adversary's malicious behaviors.
If the adversary tampers a codeword of a non-malleable code, its
decoding results in either the original message or an independent
message of the original one. Thus, non-malleable codes are
applicable to situations where error-detection and error-correction
are impossible. For example, they provides a security guarantee
against adversaries who can overwrite encoded messages.

We suppose that the adversary tampers a codeword $C$ by
applying a function $f$ to $C$. We consider the situation
where a message $s\in\{0,1\}^k$ is randomly encoded and the
encoded message is tampered by $f$. We denote the resulting 
corrupted codeword by a random variable ${\it Tamper}^f_s$.
For the non-malleability, it is desirable that,
for any $s,s'\in\{0,1\}^k$, the random variables
${\it Tamper}^f_s$ and ${\it Tamper}^f_{s'}$ are almost
identical to each other. But, it may happen 
that the decoding result $\tilde{s}$ 
for a tampered codeword coincides with the original
message $s$. In this case, it is clear that
${\it Tamper}^f_s$ is dependent on $s$.
Thus, we consider a probability distribution $D_f$ whose support
includes $\tilde{s}$ and a special symbol ${\it same}^\ast$.
By using the above probability distribution, the notion of non-malleability
codes can be defined. A code is non-malleable 
if there exists a probability distribution $D_f$ such that,
for any $s\in\{0,1\}^k$,
the following two probability distributions are statistically 
indistinguishable: (1) the induced probability distribution
from ${\it Tamper}^f_s$ and (2)
the probability distribution which is the identical to $D_f$ but
if ${\it same}^\ast$ appears then we replace it with $s$.

%

In general, there is no non-malleablde code for any tampering functions.
In \cite{1}, Dziembowski et al.\ consider a class of bitwise
independent tampering functions and give a construction
of non-malleable codes with respect to the class of 
bitwise independent tampering functions.
Faust et al.\ \cite{9} provide efficient non-malleable codes
with respect to tampering functions
which can be computed by poly-size circuits.
Chandran et al.\ \cite{7} consider block-wise tampering and show
the impossibility of non-malleable codes with respect to
block-wise tampering in the information theoretic setting. 
They also give a construction of
non-malleable codes with respect to
block-wise tampering from the viewpoint of the 
computational complexity theory.
Aggarwal et al.\ \cite{14} consider more possibility of
computational non-malleable codes.
In the literature (e.g., \cite{10,11,12,13}),
several tampering models are proposed and 
connections to other research areas such as randomness extractors 
and locally decodable codes are discussed.

In this paper, we extend bitwise independent tampering
to ``affine'' tampering, where
the adversary can falsify codewords bitwise independently
or replace some bit with the value obtained by applying an affine map
over a limited number of bits. 
We prove that the non-malleable codes with respect to bitwise 
independent tampering, provided by Dziembowski et al.\ \cite{1},
are still non-malleable with respect to the affine tampering
in the information theoretic setting.


\section{Notations}
Let $g$ be a randomized function
and $g(x;r)$ be the functional value on input $x$
which can be computed with supplimentary randomness $r$.
If we do not have to specify the randomness $r$, we
denote it by $g(x)$.
If $D$ is a probability distribution,
$d\leftarrow D$ means that a value $d$ is chosen
according to the probability distribution $D$.
For a finite set $B$, $|B|$ denote the number of elements in $B$.
For an $n$-bit string $x\in\{0,1\}^n$,
$w_H(x)$ denotes the Hamming weight of $x$.
For two strings $x$ and $x'$ of equal length,
$d_H(x,x')\stackrel{\rm def}{=}w_H(x,x')$
denotes the Hamming distance between $x$ and $x'$.
${\it SD}(X_0,X_1) \stackrel{\rm def}{=}
\frac{1}{2}\sum_{x\in X}|P_{X_0}(x)-P_{X_1}(x)|$
denotes the statistical distance between two probability distributions
$X_0$ and $X_1$ of the same support.
If ${\it SD}(X_0,X_1)$ is negligibly small
for two probability distributions $X_0$ and $X_1$,
we say that $X_0$ and $X_1$ are statistically indistinguishable
and write $X_0\approx X_1$. If
${\it SD}(X_0,X_1)=0$, we write $X_0 =X_1$.

\section{Previous Results}
In this section, we review the previous results by
Dziembowski et al.\ in \cite{1}.

\begin{definition}(Coding Scheme) 
A coding scheme is a pair of two functions $(Enc,Dec)$,
where $Enc: \{0,1\}^k \rightarrow \{0,1\}^n$
is a (randomized) encoding function and
$Dec : \{0,1\}^n \rightarrow \{0,1\}^k \cup \{\bot\}$
is a deterministic decoding function satisfying
that $\Pr[Dec(Enc(s)) = s] = 1$ for every $s\in\{0,1\}^k$.
\end{definition}

The desired property for non-malleable codes is
discussed in Section I. We give a formal definition
of non-malleable codes below.
%

\begin{definition}(Non-malleability)
Let $F$ be a class of tampering functions
and $(Enc, Dec)$ be a coding scheme.
For each $f\in F$ and 
$s\in \{0,1\}^k$,
define a random variable as follows:
\[
{\it Tamper}^f_s 
\stackrel{\rm def}{=}
\left\{\begin{array}{c}
c\leftarrow Enc(s);\\
\tilde{c}\leftarrow f(c);\\
\tilde{s}\leftarrow Dec(\tilde{c});\\
\mbox{\rm Output}~\tilde{s}.
\end{array}\right\}. \]
The randomness of ${\it Tamper}^f_s$
comes from the randomness to compute the encoding function $Enc$.
If, for each $f\in F$ and for each $s$, there exists a universal
probability distribution $D_f$ over
$\{0,1\}^k \cup \{\bot, {\it same}^\ast\}$
such that
\[
{\it Tamper}^f_s \approx
\left\{\begin{array}{c}
\tilde{s}\leftarrow D_f;\\
\mbox{If}~\tilde{s}= {\it same}^\ast~\mbox{then output}~s;\\
\mbox{Otherwise, output}~\tilde{s}.
\end{array}\right\}
\]
then we say that $(Enc,Dec)$ is non-malleable with respect to $F$.
If the statistical distance in the above is bounded by $\varepsilon$,
we say the non-malleable code $(Enc,Dec)$ is $\varepsilon$-secure.
\end{definition}

Dziembowski et al.\ \cite{1} showed a non-malleable code
against the adversary who can tamper codewords bitwise independently.
Their construction is just a combination of
algebraic manipulation detection (AMD) codes by Cramer et al.\ \cite{3}
and a linear error-correction secret sharing scheme \cite{1}.

\begin{definition}(AMD codes \cite{3}) 
Let $(A, V)$ be a coding scheme, 
where $A : \{0, 1\}^k \rightarrow \{0, 1\}^n$ is an 
encoding function and $V$ is a decoding function.
If, for some $\rho$,
for every $s\in \{0, 1\}^k$ and for every 
$\Delta\in \{0, 1\}^n\setminus \{0^n\}$,
$\Pr[V(A(m) + \Delta)\ne \bot] \le\rho$,
then we say that $(A,V)$ is an algebraic manipulation detection (AMD)
coding scheme of $\rho$-security.
\end{definition}
\begin{definition}(LECSS scheme \cite{1}) 
Let $(E,D)$ be a coding scheme.
Suppose that $(E,D)$ satisfies the following three properties:
\begin{quote}
\underline{{\em Linearity}}:\\
For every $c\in\{0,1\}^n$ such that
$D(c)\ne\bot$ and for every
$\Delta\in \{0,1\}^n$, we have the following:
\[ D(c + \Delta) = \left\{
\begin{array}{ll}
\bot & \mbox{if}~D(\Delta) =\bot,\\
D(c) + D(\Delta) & \mbox{otherwise.}~
\end{array}\right.
\]
\underline{{\em Distance $d$}}:\\
For every $\tilde{c}\in \{0,1\}^n\setminus\{0^n\}$ 
whose Hamming weight is less than $d$,
we have $D(\tilde{c}) =\bot$.\\
\underline{{\em Secrecy $t$}}: \\
For any $s$, let
$C = (C_1,\ldots, C_n) = Enc(s)$ be a random variable,
where $C_i$ is the $i$-th bit of $C$.
Then $\{C_i\}_{1\le i\le n}$ are $t$-wise independent.
Each (marginal) $C_i$ is the uniform distribution over $\{0,1\}$.
\end{quote}
Then we say that $(E,D)$ is a $(t,d)$-linear error-correction secret-sharing
(LECSS) scheme.
\end{definition}

Bitwise independent tampering can be described as
\[ f(c_1,\ldots,c_n) = (f_1(c_1),\ldots,f_n(c_n)), \]
where each $f_i$ is 
\begin{itemize}
\item the bit-flipping function
(i.e., $f_i(b) = 1\oplus b$), 
\item
the identity function (i.e., $f_i(b)=b$), 
\item 
the 0-constant function (i.e., $f_i(b)=0$), or
\item
the 1-constant function (i.e., $f_b(1)=1$).
\end{itemize}
We denote the class of bitwise independent tampering functions
by $F_{\it BIT}$. That is,
\[ F_{\it BIT} = \left\{f=(f_1,\ldots, f_n) : f_i \in \left\{\begin{array}{@{}l@{}}
\mbox{bit-flipping, identity},\\
\mbox{0-constant, 1-constant}\end{array}\right\}
\right\}. \]


\begin{theorem}(\cite{1})\label{thm:bwit}
Suppose that $(E, D)$ is a $(d,t)$-LECSS scheme where $d>n/4$
and $(A,V)$ is a $\rho$-secure AMD coding scheme. By using these schemes,
we define a coding scheme $(End,Dec)$ as follows:
\begin{eqnarray*}
Enc(s) &=& E(A(s));\\
Dec(c) &=&
\left\{\begin{array}{ll}
\bot & \mbox{if}~D(c) =\bot,\\
V(D(c)) & \mbox{otherwise}.
\end{array}\right.
\end{eqnarray*}
Then, $(Enc,Dec)$ is $\varepsilon$-secure
non-malleable with respect to $F_{\it BIT}$,
where $\varepsilon\le\max(\rho, 2^{-\Omega(t)})$.
\end{theorem}

\section{Main Results}
In this paper, we show that Dziembowski's non-malleable code
with respect to $F_{\it BIT}$ is also non-malleable
with respect to a class of affine tampering functions, which is a
generalization of $F_{\it BIT}$.
Informally speaking, the class of affine tampering functions
includes all the bitwise independent tampering functions and also includes
functions $f$ such that $\tilde{c}_2 = f(c_1, c_2) = c_1 \oplus c_2\oplus 1$,
where bits at some positions are altered into a sum of several bits
and some constant. Here, we define a new function:
$f_i$ is said to be $\ell$-affine if 
$f_i(b_1,\ldots,b_n)=\left(\bigoplus_{j\in B} b_j\right)\oplus b$
for some bit $b\in \{0,1\}$ and some set $B\subseteq \{1,\ldots, n\}$
such that $|B|\le \ell$.
We define a class of affine tampering functions as follows:
\begin{eqnarray*}
\lefteqn{F_{\ell\mbox{\rm -}\it AFFINE}}\\
& = &
\Bigg\{f=(f_1,\ldots, f_n) : f_i \in \left\{\begin{array}{@{}l@{}}
\mbox{bit-flipping, identity},\\
\mbox{0-constant, 1-constant},\\
\mbox{$\ell$-affine}\end{array}\right\}\\
&& ~~~~~~~~\land \mbox{all $\ell$-affine functions 
are $\ell$-wise independent}\Bigg\},
\end{eqnarray*}
where functions $g_1,\ldots, g_k$ are said to be $\ell$-wise independent
if their functional values on the uniform random inputs are $\ell$-wise
independent.

\smallskip

{\it Remark:} For each $\ell$-affine function
$f_i(b_1,\ldots,b_n)=\left(\bigoplus_{j\in B} b_j\right)\oplus b$,
there is the correponding vector $\beta_i=(a_1,\ldots, a_n)$,
where $a_j=1$ if $j\in B$ and $a_j=0$ otherwise. 
Note that $w_H(\beta_i)\le\ell$. To choose $\ell$-wise independent
functions, we first choose vectors $\beta_1,\ldots,\beta_k$
such that ${\rm rank}[\beta_1\ \beta_2\ \cdots \ \beta_k]
\ge\min\{k,\ell\}$. From such vectors $\beta_1,\ldots,\beta_k$,
we can construct $k$ $\ell$-affine functions 
which are $\ell$-wise independent.


\begin{theorem}\label{thm:main}
Suppose that $(E,D)$ is a $(d,t)$-LECSS scheme where
$d > 3n/8$ and $(A,V)$ is a $\rho$-secure AMD coding scheme
and define a coding scheme $(Enc, Dec)$ as follows:
\begin{eqnarray*}
Enc(s) & = & E(A(s));\\
Dec(c) & = &
\left\{\begin{array}{ll}
\bot & \mbox{if}~D(c)=\bot,\\
V(D(c)) & \mbox{otherwise}.
\end{array}\right\}
\end{eqnarray*}
Then $(Enc,Dec)$ is $\varepsilon$-secure 
non-malleable with respect to $F_{t\mbox{\rm -}\it AFFINE}$, where
$\varepsilon\le \max(\rho, 2^{-\Omega(t)})$.
\end{theorem}
In the proof in \cite{1} that $(Enc,Dec)$ stated in Theorem \ref{thm:bwit}
is non-malleable with respect to $F_{\it BIT}$, $\{1,\ldots,n\}$
is partitioned into two subsets $B_1$ and $B_2$, where
$B_1=\{ i : f_i$ is either 0-constant or 1-constant$\}$
and $B_2=\{i : f_i$ is either bit-flipping or identity$\}$. 
They considered several cases with respect to $|B_1|$ and $|B_2|$ and
analyzed the security for each case.
We partition $\{1,\ldots,n\}$ into three subsets (say, 
$B_1$, $B_2$ and $B_3$)
and consider several cases with respect to $|B_1|$, $|B_2|$ and $|B_3|$.

\begin{proof}
We show that $(Enc,Dec)$ is non-malleable with respect to 
$F_{t\mbox{\rm -}\it AFFINE}$
and its security $\varepsilon$ satisfies
\[ \varepsilon \le \max \left(
\rho,\frac{1}{2^t}+\left(\frac{t}{n(d/n-3/8)}\right)^{t/2}\right)
\]
for any even $t>6$.
We let $f=(f_1,\ldots,f_n)$ 
be a tampering function in $F_{t\mbox{\rm -}\it AFFINE}$
and define a universal distribution $D_f$ for showing that
$(Enc,Dec)$ is non-malleable with respect to $F_{t\mbox{\rm -}\it AFFINE}$.

For any message $s\in \{0,1\}^k$, we consider
several probability distributions and use the following notations
$C^s := Enc(s)$,
$\tilde{C}^s := f(C^s)$, $\Delta^s := \tilde{C}^s - C^s$,
$\tilde{S}^s := Dec(\tilde{C}^s)$.
$C^s_i$, $\tilde{C}^s_i$ and $\Delta^s_i$ 
for each $i\in\{1,\ldots, n\}$ denote
the $i$-th bit of $C^s$, $\tilde{C}^s$ and $\Delta^s$, respectively.
We partition $i\in\{1,\ldots, n\}$ into three subsets $B_1$, $B_2$ and $B_3$
as follows:
$B_1=\{ i : f_i$ is 0-constant or 1-constant$\}$,
$B_2=\{ i : f_i$ is bit-flipping or identity$\}$
and $B_3=\{ i : f_i$ is $t$-affine$\}$.
We let $p=|B_1|$, $q=|B_2|$ and $r=|B_3|$, which satisfy $p+q+r=n$.
We define a probability distribution ${\it Patch}(D_f,s)$ 
as follows: First, sample $\tilde{s}$ as
$\tilde{s}\leftarrow D_f$.
If $\tilde{s}={\it same}^\ast$ then output $s$ instead of
${\it same}^\ast$. Otherwise, output $\tilde{s}$ as it is.
We will construct $D_f$ such that,
for any $s$, $SD(\tilde{S}^s,{\it Patch}(D_f,s))\le \varepsilon$.
Before discussing each case, we need some useful property:

{\it Fact:} 
If $i\in B_3$ then $\tilde{C}^s_i$ is the uniform distribution
over $\{0,1\}$ and the joint distribution
$\{ \tilde{C}^s_i\}_{i\in B_3}$ 
is $t$-wise independent because of $t$-secrecy of the
LECSS scheme and $t$-wise independence of affine functions
in $F_{t\mbox{\rm -}{\it AFFINE}}$ for any $s$.
So is 
$\{\Delta^s_i = \tilde{C}^s_i - C^s_i \}_{i\in B_3}$.

\subsubsection*{Case 1} $p\le t-r$\\
We show that $\Delta^s$ for each $s$ is identical 
to $\Delta^{s'}$ for any other $s'$.
\begin{itemize}
\item If $f_i$ is the identity function, then we have $\Delta^s_i = 0$. 
If $f_i$ is bit-flipping, then we have $\Delta^s_i = 1$.
\item If $i\in B_1\cup B_3$ then $\Delta^s_i$ is the uniform distribution
over $\{0,1\}$, since $|B_1\cup B_3| = |B_1|+|B_3| = p+r \le t$
imply that $\{C^s_i\}_{i\in B_1\cup B_3}$ is $t$-wise independent.
Thus, we have
$\{\Delta^s_i = \tilde{C}^s_i - C^s_i\}_{i\in B_1\cup B_3}$
is the uniform distribution regardless of $s$.
\end{itemize}
Therefore, there exists a universal probability distribution $\Delta$
such that $\Delta = \Delta^s$ for any $s$ and we have
\begin{eqnarray}
\tilde{S}^s & = & Dec(\tilde{C}^s)\nonumber\\
&=& V(D(C^s + \Delta^s))\nonumber\\
&=& V(D(C^s) + D(\Delta^s))\label{eq:1}\\
&=& V(A(s) + D(\Delta^s))\nonumber\\
&=& V(A(s) + D(\Delta)),\nonumber
\end{eqnarray}
where (\ref{eq:1}) is by the linearity of the LECSS scheme.
\begin{enumerate}
\item If $D(\Delta)\ne 0$ then the security of
AMD codes imply that $\Pr[\tilde{S}^s =\bot] \ge 1 -\rho$.
\item If $D(\Delta) = 0$ then we have $\Pr[\tilde{S}^s = s] = 1$.
\end{enumerate}
From 1) and 2), we define $D_f$ as follows:
First, sample $\delta$ as $\delta\leftarrow \Delta$.
If $D(\delta) = 0$ then output ${\it same}^\ast$. Otherwise,
output $\bot$. Then, we have
$SD(\tilde{S}^s,{\it Patch}(D_f, s))\le \rho$ for any $s$.
This completes the proof in Case 1.

\subsubsection*{Case 2} $p\ge n-t$\\
In this case, we show that
$\tilde{C}^s$ for each $s$ is identical to 
$\tilde{C}^{s'}$ for any other $s'$.
\begin{itemize}
\item If $f_i$ is 0-constant,
then $\tilde{C}^s_i=0$. If $f_i$ is 1-constant,
then $\tilde{C}^s_i=1$.
\item For any $i \in B_2 \cup B_3$, $\tilde{C}^s_i$
is the uniform distribution over $\{0,1\}$, 
since $p\ge n-t$ implies that $|B_2 \cup B_3| =|B_2| + |B_3| = q + r \le t$.
Thus, we can say that $\{C^s_i\}_{i\in B_2\cup B_3}$ 
is the uniform distribution
and $\{\tilde{C}^s_i = f(C^s_i)\}_{i\in B_2\cup B_3}$ 
are independent uniform distributions for any $s$.
\end{itemize}
Furthermore, there exists a universal distribution
$\tilde{C}$ such that $\tilde{C} = \tilde{C}^s$ for any $s$
and we have 
$\tilde{S}^s = Dec(\tilde{C}^s) = Dec(\tilde{C})$.
We define the distribution $D_f$ which samples $\tilde{C}$ as above
and computes $Dec(\tilde{C})$.
This implies that
$SD(\tilde{S}^s,{\it Patch}(D_f,s)) = SD(\tilde{S}^s,D_f) = 0$
for any $s$. This completes the proof in Case 2.

\subsubsection*{Case 3}  $t-r < p \le (n-r)/2$\\
In this case, we show that a probability 
distribution that always outputs $\bot$ is
a universal distribution $D_f$.
Since, for any $s$,
\begin{eqnarray*}
\Pr[\tilde{S}^s\ne\bot] & = & \Pr[Dec(\tilde{C}^s)\ne\bot]\\
& = & \Pr[D(\Delta^s)\ne\bot],
\end{eqnarray*}
it suffices to show that $\Pr[D(\Delta^s)\ne\bot]$ is small.
$\{\Delta^s_i\}_{i\in B_2}$ is fixed to a constant by $f$
(if $f_i$ is the identity function then $\Delta^s_i$ is fixed to $0$ and
if $f_i$ is bit-flipping then $\Delta^s_i$ is fixed to $1$)
Let $\delta^\ast\in \{0,1\}^n$ be any value which is consistent
with the fixed bits of $\Delta$ so that
$\{\Delta^s_i = \delta^s_i \}_{i\in B_2}$
and for which $D(\delta^*) \ne\bot$.
If no such value exists then we are done
since $D(\Delta^s) =\bot$ with probability 1.
So let us assume that some such value exists.
Since $t < p+r \le (n+r)/2$,
$\{\Delta^s_i\}_{i\in B_1\cup B_3}$ are $t$-wise independent
uniform distributions and we have $\Pr[\Delta^s =\delta^\ast]\le
1/2^t$.
On the other hand, we show that $d_H(\Delta^s,\delta^\ast)$ is not so large.
The expected value of the Hamming distance between $\Delta^s$ and $\delta^\ast$
satisfies the following.
\begin{eqnarray}
{\bf E}[d_H(\Delta^s, \delta^\ast)] 
	&=& {\bf E}[\sum_{i=1}^n d_H(\Delta^s_i,\delta^\ast_i)]\nonumber\\
&=& {\bf E}[\sum_{i\in B_1\cup B_3} d_H(\Delta^s_i,\delta^\ast_i)] \label{eq:2}\\
&=& \sum_{i\in B_1\cup B_3} {\bf E}[d_H(\Delta^s_i,\delta^\ast_i)] \label{eq:3}\\
&=& \frac{p+r}{2}.\label{eq:4}
\end{eqnarray}
In the above, (\ref{eq:2}) holds since
$\Delta^s = \delta^\ast$ for $i\in B_2$ 
and thus $\{d_H(\Delta^s_i,\delta^\ast_i)\}_{i\in B_2} =0$.
(\ref{eq:3}) is by the linearity of the expectation.
For (\ref{eq:4}), since
$\Delta^s_i$ are independent for $i\in B_1\cup B_3$,
we consider the probability that
$d_H(\Delta^s,\delta^\ast) =\sum_{i\in B_1\cup B_3} d_H(\Delta^s_i,\delta^\ast_i)$
is larger than $d$.
Since 
$\{d_H(\Delta^s_i,\delta^\ast_i)\}_{i\in B_1\cup B_3}$ 
are $t$-wise independent, we can apply
a Chernoff-Hoeffding tail bound as in \cite{5,6}.
Thus, we have
\begin{eqnarray}
\lefteqn{\Pr[d_H(\Delta^s,\delta^\ast) \ge d]}\nonumber\\
& \le &
\Pr\left[
\left| d_H(\Delta^s_i, \delta^\ast_i) - \frac{p+r}{2}\right|
\ge d - \frac{p+r}{2} \right] \nonumber\\
& \le & \left(\frac{nt}{(d-\frac{p+r}{2})^2} \right)^{t/2} \label{eq:5}\\
& \le & \left(\frac{nt}{(d-\frac{n+r}{4})^2} \right)^{t/2} \nonumber\\
& \le & \left(\frac{t}{n(\frac{d}{n}-\frac{3}{8})^2} \right)^{t/2}.\label{eq:6}
\end{eqnarray}
In the above, (\ref{eq:5}) follows from Lemma 2.2 in \cite{5} 
by Bellare and Rompel.
For (\ref{eq:6}), we use $r< n/2$ since $r \le t$. Hence, we have
\begin{eqnarray*}
\lefteqn{\Pr[D(\Delta^s)\ne \bot]}\\
& \le &
\Pr[\Delta^s = \delta^\ast \lor d_H(\Delta^s,\delta^\ast)\ge d ]\\
&\le & \frac{1}{2^t} + \left(\frac{t}{n(\frac{d}{n}-\frac{3}{8})^2}\right)^{t/2}
\end{eqnarray*}
and this completes the proof in Case 3.

\subsubsection*{Case 4} $(n-r)/2 < p \le n-t$\\
In this case, 
we show that a probability 
distribution that always outputs $\bot$ suffices
for a universal distribution $D_f$.
To this end, we show that the probability that
$\Pr[\tilde{S}^s \ne\bot] = \Pr[D(\tilde{C}^s)\ne\bot]$ 
is small for any $s$.
Since $(n-r)/2 < p \le n-t$, we $t < q+r < (n+r)/2$ and thus
$\{\tilde{C}^s_i\}_{i\in B_1}$ are fixed by $f$.
Let $\tilde{c}^\ast\in\{0,1\}^n$ be any value which is
consistent with the fixed portion of $\tilde{C}^s$ 
so that $\{\tilde{C}^s_i = \tilde{c}^s_i\}_{i\in B_1}$.
If no such value exist then we are done.
Otherwise, we can use the similar discussion as in Case 3
and we have
\begin{eqnarray*}
\lefteqn{\Pr[d_H(\tilde{C}^s,\tilde{c}^\ast)\ge d]}\\
&\le& \Pr[\tilde{C}^s = \tilde{c}^\ast \lor d_H(\tilde{C}^s,\tilde{c}^\ast)
\ge d]\\
&\le& \frac{1}{2^t} + \left(\frac{t}{n(\frac{d}{n}-\frac{3}{8})^2}\right)^{t/2}.
\end{eqnarray*}
This complete the proof in Case 4.

\medskip

For any cases of $p,q,r$, we have completed the proof.
Thus, we can say that Theorem \ref{thm:main} holds.
\end{proof}

\medskip

{\it Remark:} 
In Theorem \ref{thm:main}, we use a $(d,t)$-LECSS code where
$d > 3n/8$. This requires that an LECSS code for
non-malleability with respect to $F_{t\mbox{\rm -}\it AFFINE}$
must be better than ones 
with respect to $F_{\it BIT}$. Chen et al.\ \cite{4} have shown the
existence such LECSS codes.

\section{Concluding Remarks}
We have extended the bitwise independent tampering 
to the affine tampering and shown that the non-malleable
codes in \cite{1} with respect to the bitwise independent tampering
is also non-malleable with respect to the affine tampering.
Our tampering model for the affine tampering may be a bit artificial
because of some technical reason. As mentioned, the property of 
being ``affine'' is useful to construct $\ell$-wise independent functions.
But, this does not rule out the possibility to
construct $\ell$-wise independent functions from non-affine tampering
functions. Thus, in future, we may find a wider class of tampering
functions for which there exists a non-malleablde coding scheme.


\end{document}